\documentclass[aps,prl,reprint,superscriptaddress,amsmath,amssymb,floatfix,longbibliography]{revtex4-1}

\usepackage{graphicx}
\usepackage{dcolumn}
\usepackage{bm}
\usepackage{color}
\usepackage[utf8]{inputenc}
\usepackage{physics}
\usepackage{mathtools}
\usepackage{soul}

\renewcommand{\op}[1]{\hat{#1}}

\begin{document}
\title{Numerical Verification of 
Fluctuation Dissipation Theorem for Isolated Quantum Systems}
\author{Jae Dong Noh}
\affiliation{Department of Physics, University of Seoul, Seoul 02504, Korea}
\author{Takahiro Sagawa}
\affiliation{Department of Applied Physics, The University of Tokyo, 7-3-1
Hongo, Byunkyo-ku, Tokyo 113-8656, Japan}
\author{Joonhyun Yeo}
\affiliation{Department of Physics, Konkuk University, Seoul 05029, Korea}

\date{\today}

\begin{abstract}
  The fluctuation dissipation theorem~(FDT) is a hallmark of thermal equilibrium
  systems in the Gibbs state. We address the question whether the FDT is
  obeyed by isolated quantum systems in an energy eigenstate.
  In the framework of the eigenstate thermalization 
  hypothesis, we derive the formal expression for two-time correlation
  functions in the energy eigenstates or in the diagonal ensemble. 
  They satisfy the Kubo-Martin-Schwinger condition, which is the sufficient and
  necessary condition for the FDT, in the infinite system size limit. 
  We also obtain the finite size correction to the FDT for finite-sized
  systems. With extensive numerical works for 
  the XXZ spin chain model, we confirm our theory for the FDT and the 
  finite size correction. 
  Our results can serve as a guide line for an experimental study of the FDT 
  on a finite-sized system. 
\end{abstract}

\maketitle

{\em Introduction---}
It is a fascinating question to ask when and how isolated
quantum many body systems approach the thermal equilibrium state. 
Recent advances in experimental techniques with ultracold atoms boost
research interests, theoretical and experimental, in quantum
thermalization~\cite{Kinoshita:2006bg,Trotzky:2012iu,Langen:2015do,Kaufman:2016jm,Gross:2017do,Tang:2018dq}.
The relaxation of an isolated quantum system 
into a stationary state has been proved in a broad range of systems and initial
states~\cite{Deutsch:1991ju,Reimann:2008hq,Short:2011fq,Wilming:2019fj}.
Thermalization furthermore requires that this stationary state is indistinguishable from
the equilibrium microcanonical state. 

It is the eigenstate thermalization hypothesis~(ETH) that makes a link between 
the stationary state and the Gibbs state.
The ETH is an assumption for matrix elements of local observables in the
Hamiltonian eigenstate
basis~\cite{Srednicki:1996kn,DAlessio:2016gr,Deutsch:2018fy}. Under the ETH,
expectation values of observables in
the Hamiltonian eigenstates coincide with the microcanonical
ensemble averages.
The ETH has been tested extensively, and is believed to hold in 
nonintegrable systems~\cite{Rigol:2008bf,Kim:2014kw,Yoshizawa:2018js}. 
It is confirmed that
matrix elements of observables display the statistical properties postulated by the
ETH~\cite{Rigol:2009ew,Steinigeweg:2013dc,Mondaini:2016dn,Mondaini:2017jg,Mierzejewski:2020jm}. 
Thermalization after quantum quench~\cite{Santos:2011ip,Mallayya:2018ki} and 
thermodynamic processes such as the Joule expansion are also understood 
well in the framework of the ETH~\cite{Camalet:2008fg,Noh:2019gx}.

The fluctuation dissipation theorem~(FDT), which provides the universal relation between
the response~(or dissipation) and the correlation~(or fluctuation), is another hallmark of thermal
equilibrium states~\cite{Kubo:1985bs,Mazenko:2006vd}. 
As equilibrium dynamics is
characterized by the detailed balance, the dynamic response function 
and the correlation function of equilibrium systems are not independent 
but tightly linked to each other. 
The FDT has been used to 
distinguish equilibrium and nonequilibrium 
dynamics~\cite{Kurchan:2005kh} and to measure the temperature of 
microscopic quantum systems~\cite{Gemelke:2009ja,Mehboudi:2019hz}. 

There have been growing number of studies on the connection between the quantum thermalization
and the fluctuation and dissipation in isolated quantum systems.
{Foini {\em et al.}~\cite{Foini:2012gq,Foini:2017dl} investigated the
relaxation dynamics of an integrable quantum system which is
nonthermal.}
{Essler {\em et al.}~\cite{Essler:2012hk} proposed an argument connecting
static and dynamic correlations based on the Lieb-Robinson 
bound~\cite{Lieb:1972kt}.} Srednicki~\cite{Srednicki:1999bo} studied 
a version of FDT for isolated quantum systems which involves
the correlation of the {\em expectation values} of observables. 
This in fact corresponds to the classical limit of the full quantum mechanical FDT
and has been further investigated in Refs.~\cite{Khatami:2013kf,Nation:2019kr}.
D'Alessio {\em et al.}~\cite{DAlessio:2016gr} demonstrated that the FDT with a
single observable is consistent with the ETH. The FDT with two different 
observables requires an assumption on the behavior of the random variables 
arising in the ETH, which needs to be verified.

In this Letter, we present an explicit numerical verification and a comprehensive study 
of the FDT in isolated quantum systems. 
In the framework of the ETH, we derive a symmetry relation among the quantum 
mechanical two-time correlation functions, known as the
Kubo-Martin-Schwinger~(KMS) condition~\cite{Haag:1967sg,Mazenko:2006vd}.
Combining the symmetry relation and the linear response theory, 
we show that an isolated quantum system in an energy
eigenstate obeys the FDT in the {\em infinite size limit}.
Finite-sized systems violate the FDT. 
We derive the analytic expression for
the {\em finite size correction} to the FDT. When the energy uncertainty or
variance of the quantum state scales as $\Delta^2 E = O(L^{d})$ or is smaller
than that, 
the finite size correction term scales as
$O(L^{-d})$ with the system size $L$ and the spatial dimension $d$.
We verify our analytic theory with
the exact diagonalization study for the XXZ spin chain, equivalently the
hardcore boson model, in one-dimensional lattices.
We demonstrate the finite size correction 
to the FDT using the energy eigenstate.

{\it KMS Condition and FDT---}
A quantum system with Hamiltonian $\op{H}$ is in an initial state 
$\op{\rho}_i$ at time $t=t_0$. When a perturbation $\delta\op{H} = -h(t) \op{B}$ is applied, 
an expectation value of an observable $\op{A}$ at time $t>t_0$ deviates from its unperturbed value. 
According to the linear response theory~\cite{Mazenko:2006vd,Tauber:2014cj}, the deviation is given by
$\delta A(t)  = 2 i \int_{t_0}^t dt' \chi''_{AB}(t,t') h(t') +O(h^2)$
with the linear response function 
\begin{equation}\label{chi''}
	\begin{split}
\chi''_{AB}(t,t') &\equiv \frac{1}{2\hbar} \left\langle [\op{A}(t),\op{B}(t')]
\right\rangle_i \\
 &=  \frac{1}{2\hbar} \left( \bar{S}_{AB}(t,t') -
\bar{S}_{BA}(t',t) \right).
\end{split}
\end{equation}
The operators are in the Heisenberg picture with respect to the
unperturbed Hamiltonian, $\langle \rangle_i$ stands for the expectation
value in the state $\op{\rho}_i$, and $\bar{S}_{AB}(t,t') \equiv \langle
\op{A}(t)\op{B}(t')\rangle_i - \langle \op{A}(t)\rangle_i \langle
\op{B}(t')\rangle_i$ is the two-time connected correlation function.
The response function is defined for both $t\geq t'$ and $t<t'$. 
The causal response function is given by $\chi_{AB}(t,t') = 2i\Theta(t-t') 
\chi_{AB}''(t,t')$ with the Heaviside step function $\Theta(t-t')$. 

Suppose that the system is prepared in the thermal equilibrium state
characterized by the Gibbs state $\op{\rho}_i = \op{\rho}_{eq}(\beta) =
e^{-\beta \op{H}}/Z$ with inverse temperature $\beta$ and the partition
function $Z$.  
Throughout the paper, we set the Boltzmann constant $k_B=1$.
The equilibrium state is stationary so that the correlation
function and the response function depend on the time difference $t-t'$.
Furthermore, since the Boltzmann factor $e^{-\beta\op{H}}$ is equal to the
time evolution operator in the imaginary time direction, the correlation
functions obey the KMS condition~\cite{Haag:1967sg,Mazenko:2006vd} that 
\begin{equation}\label{KMS_t}
    \bar{S}_{AB,eq}(t) = \bar{S}_{BA,eq}(-t-i\beta\hbar) 
\end{equation}
or, equivalently, 
\begin{equation}\label{KMS_w}
    \bar{S}_{AB,eq}(\omega) = \bar{S}_{BA,eq}(-\omega) e^{\beta\hbar\omega}
\end{equation}
in the frequency domain. A Fourier transformation is
defined by $\bar{S}_{AB,eq}(\omega) = \int_{-\infty}^\infty dt
\bar{S}_{AB,eq}(t) e^{i\omega t}$.
Combining the KMS condition and the linear response theory, one obtains 
the celebrated quantum mechanical {\em fluctuation dissipation theorem}
\begin{equation}\label{qFDT}
    \chi_{AB,eq}''(\omega) = \frac{1-e^{-\beta\hbar\omega}}{2\hbar} \bar{S}_{AB,eq}(\omega) .
\end{equation}
In the classical limit where $\hbar\to 0$, it becomes 
\begin{equation}
    \bar{S}_{AB,eq,cl}(\omega) = \frac{2}{\beta \omega}
	\chi_{AB,eq,cl}''(\omega) .
\end{equation}
Integrating over all $\omega$, one obtains the familiar relation
\begin{equation}
  \chi_{AB,eq,cl}(\omega=0) = \beta \bar{S}_{AB,eq,cl}(t=0)
\end{equation}
between the static susceptibility and the equal time 
correlation function~\cite{Mazenko:2006vd}. 
We stress that the KMS condition in Eq.~\eqref{KMS_t} or \eqref{KMS_w} 
is the necessary and sufficient condition for the FDT in Eq.~\eqref{qFDT}
provided that the linear response theory is valid.

{\em FDT from ETH---} We address the question whether the FDT holds 
for a generic nonintegrable quantum system not necessarily in the Gibbs state.
In the framework of the ETH, we formulate the FDT with a focus on the KMS
condition. Note that the KMS condition can be examined numerically easily,
as will be shown later.
We will denote Hamiltonian eigenstates and eigenvalues as
$\{|\alpha\rangle\}$ and $\{E_\alpha\}$. 

Suppose that the initial state is given by 
$\op{\rho}_i = \sum_\alpha p_\alpha |\alpha\rangle\langle\alpha|$.
The mean energy and the energy variance are given by
$\bar{E} =  \langle \op{H}\rangle_i$ and 
$\Delta^2 E = \langle(\op{H}-\bar{E})^2\rangle_i$. 
Such a state is called the diagonal ensemble which corresponds to the
stationary state limit of a pure state $|\psi\rangle = \sum_\alpha
c_\alpha \ket{\alpha}$ with $p_\alpha = |c_\alpha|^2$~\cite{Rigol:2008bf}.
An energy eigenstate  $|\alpha_0\rangle\langle \alpha_0|$ 
is a special case with $p_\alpha = \delta_{\alpha \alpha_0}$. 
The inverse temperature $\beta$ of the initial state is determined by 
$\bar{E} = {\rm Tr}~ \op{H} \op{\rho}_{eq}(\beta)$. 
Let $\op{A}$ and $\op{B}$ be Hermitian operators for observables.
The correlation function $\bar{S}_{AB}(\omega)$ for $\omega\neq 0$ is given by
\begin{equation}\label{S_AB_formal}
\bar{S}_{AB}(\omega) = 2\pi \sum_{\alpha} \sum_{\gamma\neq \alpha} p_\alpha
    A_{\alpha\gamma} B_{\gamma\alpha}\delta(\omega-\omega_{\gamma\alpha})
\end{equation}
with $\omega_{\gamma\alpha} \equiv (E_\gamma-E_\alpha)/\hbar$ and 
$A_{\alpha\gamma}=\langle\alpha\vert\op{A}\vert\gamma\rangle$, etc..
If the initial state is the equilibrium Gibbs state 
$\op{\rho}_{eq}(\beta)$, each term in $\bar{S}_{AB}$ and $\bar{S}_{BA}$ 
has the ratio $p_\alpha/p_\gamma = e^{\beta \hbar
\omega_{\gamma\alpha}}$. Thus, the KMS condition \eqref{KMS_w} holds
identically regardless of characteristics of the operators. 
For nonthermal states, however, the KMS condition requires a specific
property of the operators.

According to the ETH, matrix elements of a Hermitian operator $\op{X}$ 
in the energy eigenstate basis has the structure 
\begin{equation}\label{eth_ansatz}
    X_{\gamma\alpha} = X(E_{\gamma\alpha}) \delta_{\gamma\alpha} +
    e^{-S(E_{\gamma\alpha})/2}
    f_X(E_{\gamma\alpha},\omega_{\gamma\alpha})
    R^X_{\gamma\alpha} ,
\end{equation}
where $E_{\gamma\alpha} = (E_\gamma+E_\alpha)/2$,
$\omega_{\gamma\alpha} =  (E_\gamma-E_\alpha)/\hbar$, 
$S(E)$ is the microcanonical ensemble entropy, $R^X$ is a random 
matrix in the Gaussian unitary ensemble, 
and $X(E)$ and $f_X(E,\omega) = f_X(E,-\omega)^*$ 
are smooth functions~\cite{Srednicki:1996kn,DAlessio:2016gr}.
Using \eqref{eth_ansatz} for $\op{A}$ and $\op{B}$, it is straightforward to
obtain that~\footnote{See the Supplemental Materials.}
\begin{equation}\label{S_AB_result}
	\begin{split}
    \bar{S}_{AB}(\omega) =& 2\pi \exp\left[\frac{1}{2}\beta\hbar\omega
    + \mathcal{Y}_{AB}(\bar{E},\omega)\right]  \\
    &\times
    f_A(\bar{E},-\omega) f_B(E,\omega) \mathcal{R}_{AB}(\bar{E},\omega) ,
    \end{split}
\end{equation}
where 
$\mathcal{R}_{AB}(E,\omega)$, called an overlap function, will be explained
below and $\mathcal{Y}_{AB}(E,\omega)$ is a finite-size correction term. It
consists of an intrinsic term $\mathcal{Y}^{(1)}= O(L^{-d})$ 
and an extrinsic term $\mathcal{Y}^{(2)} = O(\Delta^2
E/L^{2d})$ arising from the energy uncertainty. 
When the energy variance scales as $\Delta^2 E = O(L^d)$ or is
smaller than that, we have
\begin{equation}
	\mathcal{Y}_{AB}(E,\omega)= O(L^{-d}) .
\end{equation}
The detailed derivation and the explicit expression of $\mathcal{Y}$ are presented 
in Supplemental Material~\cite{Note1}.

Matrix elements $R^A_{\alpha\gamma}$ and $R^B_{\gamma\alpha}$ are
random variables, so are their products $R^A_{\alpha\gamma}
R^B_{\gamma\alpha}$. 
The overlap function $\mathcal{R}_{AB}(E,\omega)$ 
is defined as the mean value of $R^A_{\alpha\gamma} R^B_{\gamma\alpha}$ among
all pairs of eigenstates such that
$(E_\gamma+E_\alpha)/2 = E$ and $(E_\gamma-E_\alpha)/\hbar =
\omega$ within the infinitesimal range~\cite{Note1}:
\begin{equation}\label{R_def}
  R^A_{\alpha\gamma}R^B_{\gamma\alpha}
  = \mathcal{R}_{AB}(E=E_{\gamma\alpha},\omega=\omega_{\gamma\alpha}) 
  + \eta^{AB}_{\gamma\alpha}
\end{equation}
with a random variable $\eta^{AB}_{\gamma\alpha}$ of zero mean.
The overlap function is similar but slightly different from the noise kernel
of Ref.~\cite{DAlessio:2016gr}.
When $\op{A}=\op{B}$, it is trivial that
$\mathcal{R}_{AA}(E,\omega)=1$.
The overlap function reflects a quantum mechanical correlation between two
observables, and is a crucial ingredient for the FDT~\cite{DAlessio:2016gr}. 
Its existence will be verified numerically shortly.
For Hermitian operators, 
$\mathcal{R}_{AB}(E,\omega) = \mathcal{R}_{AB}(E,-\omega)^* = 
\mathcal{R}_{BA}(E,-\omega)$. 

The KMS condition, hence the FDT, can be examined with 
an indicator function
\begin{equation}\label{g_def}
g_{AB}(\omega) = \frac{1}{\hbar\omega} \ln\left[
\frac{\bar{S}_{AB}(\omega)}{\bar{S}_{BA}(-\omega)}\right].
\end{equation}
When the FDT is valid, the indicator function is independent of $\omega$ and
equal to the inverse temperature $\beta$.
The analytic result \eqref{S_AB_result} leads to 
$g_{AB}(\omega) = \beta + \delta\beta_{AB}(\omega)$ with a deviation from
the FDT given by
\begin{equation}\label{beta_fsc}
\delta\beta_{AB}(\omega) = \frac{1}{\hbar\omega}\left( \mathcal{Y}_{AB}(\bar{E},\omega) -
\mathcal{Y}_{BA}(\bar{E},-\omega) \right).
\end{equation}
It vanishes as $\delta\beta_{AB} = O(L^{-d})$ for $\Delta^2 E
\leq O(L^d)$ with the system size. 
Therefore, we conclude that the ETH system obeys the KMS condition, hence
the FDT, in the thermodynamic limit. 

{\em Numerical test of the FDT and finite size effect---} 
We perform the numerical analysis 
to verify the FDT and the finite size effect for isolated quantum systems.
The indicator function $g_{AB}(\omega)$ in \eqref{g_def} is a useful
measure. If the FDT is valid, it should be a constant equal to the 
inverse temperature. 
In this work, we focus on the {\em energy eigenstate} initial state 
with $\Delta^2E=0$. 

We study the spin-1/2 XXZ spin model with nearest and next nearest neighbor 
couplings in the one-dimensional chain of $L$ sites under the periodic 
boundary condition~\cite{Yoshizawa:2018js,Kim:2014kw}.
The Hamiltonian is given by $\op{H} = \frac{1}{1+\lambda} \sum_{l=1}^L 
\left[ \op{h}_{l,l+1} + \lambda \op{h}_{l,l+2}\right]$ with $\op{h}_{l,m} = 
-J \left( \op{\sigma}_l^+\op{\sigma}_m^- +\op{\sigma}_l^- \op{\sigma}_m^+ + 
\frac{\Delta}{2} \op{\sigma}_{l}^z \op{\sigma}_{m}^z \right)$ with the Pauli matrices.
The system is nonintegrable with nonzero $\lambda$. 
We focus on the subspace in which states have zero magnetization and are invariant
under the translation, the spatial inversion, and the spin
reversal. 
The Hamiltonian is diagonalized exactly numerically~(see e.g., Ref.~\cite{Jung:2020ia}). 
We set $\hbar=1$ and fix $J=1$, $\Delta=1/2$, and $\lambda=1$ 
in numerical calculations.

\begin{figure}[t]
  \includegraphics[width=\columnwidth]{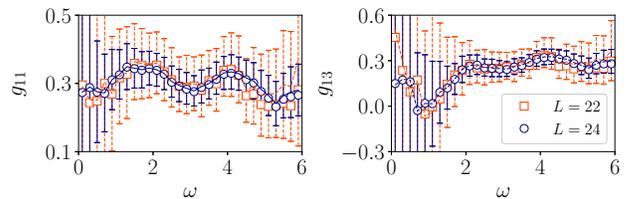}
  \caption{Mean values~(lines) and standard deviations~(error bars) of
  $g_{ij}(\omega)$'s for $\beta_{\mathrm T}=0.3$ at $L=22$~(dashed) and $24$~(solid). 
    The indicator function from the averaged correlation
functions is plotted with symbols.}
  \label{fig:1}
\end{figure}

We choose an energy eigenstate $\ket{\alpha_{\textrm{T}}}$ whose inverse
temperature is closest to a target value $\beta_{\textrm{T}}$, and evaluate a
coarse-grained $\bar{S}_{AB}(\omega)$ for a set of discretized $\omega$'s in
unit of $\Delta\omega=0.2$~\cite{Note1}.
The indicator function fluctuates from eigenstate to eigenstate.
Figure~\ref{fig:1} exemplifies the fluctuations of $g_{ij}(\omega)$ 
for operators $\hat{A}=\hat{O}_{i}$ and $\hat{B}=\hat{O}_j$~(see next
paragraphs for $\hat{O}_i$).
It shows the mean value and the standard deviation of
$g_{ij}(\omega)$ among eigenstates $\ket{\alpha}$'s within a window 
$|E_\alpha-E_{\mathrm T}| \leq \frac{\Delta\omega}{2}$ with $\beta_{\mathrm T}=0.3$.
The standard deviation decreases by a factor $\sim 2$ as $L$ increases from
$22$ to $24$, which suggests that the eigenstate-to-eigenstate fluctuations
vanish in the thermodynamic limit. 
Moreover, the mean value is in perfect agreement with 
the indicator function obtained from the correlation functions averaged within 
the window. Based on these observations, we will focus on the indicator function 
calculated from the averaged correlation functions~\cite{Note1}.

\begin{figure}[t]
  \includegraphics[width=\columnwidth]{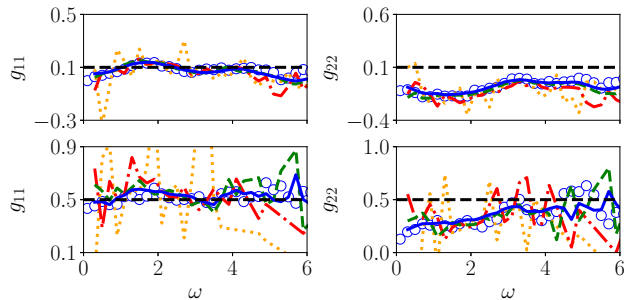}
  \caption{FDT indicator functions $g_{11}$ for $\op{A}=\op{B}=\op{O}_1$~(left column) and
  $g_{22}$ for $\op{O}_2$~(right column). The target inverse temperatures
  are $\beta_{\textrm{T}}$ = 0.1~(top) and 0.5~(bottom), which are
  marked with horizontal lines. The lattice sizes are $L=18$~(dotted),
  20~(dashed dotted), 22~(dashed), and
  24~(solid). Predictions from the leading order finite size effect for $L=24$ 
  are drawn with symbols.}
  \label{fig:2}
\end{figure}

\begin{figure}[t]
	\includegraphics[width=\columnwidth]{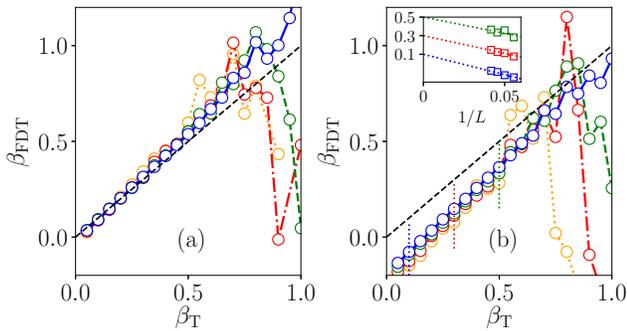}
	\caption{$\beta_\textrm{FDT}$ vs $\beta_\textrm{T}$ for
    $\op{A}=\op{B} = \op{O}_1$ in (a) and $\op{O}_2$ in (b). Inset
    illustrates the $1/L$ dependence of $\beta_{\mathrm{FDT}}$ at
    $\beta_{\mathrm T}=0.1, 0.3, 0.5$ with lines as guides to eyes.}
	\label{fig:3}
\end{figure}

Firstly, we present the numerical results for single-operator 
cases $\op{A}=\op{B} = \hat{O}_1 \equiv \sum_l \op{\sigma}^z_l
\op{\sigma}^z_{l+1}$~[nearest neighbor interaction energy] and 
$\hat{O}_2 \equiv \frac{1}{L}\sum_{l,m} \op{\sigma}_l^+ \op{\sigma}_m^-$~[zero
momentum distribution]. 
The indicator functions are shown in Fig.~\ref{fig:2}. 
There are noisy fluctuations, which weaken as $L$
increases.  For a quantitative analysis, we measure the mean value of the
indicator function $g(\omega)$ in the interval $1<\omega<5$. 
It is denoted as $\beta_{\textrm{FDT}}$, and  
plotted as a function of $\beta_\textrm{T}$ in Fig.~\ref{fig:3}.  
For the operator $\op{O}_1$, the plot tends to align
with the line $y=x$ as $L$ increases. This may be regarded as a numerical
evidence for the FDT. However, 
the systematic $\omega$ dependence of $g(\omega)$ in Fig.~\ref{fig:2} and 
a rather conspicuous deviation for $\op{O}_2$ in Fig.~\ref{fig:3}(b)
may make the general validity of the FDT questionable. 

\begin{figure}[htp]
	\includegraphics[width=\columnwidth]{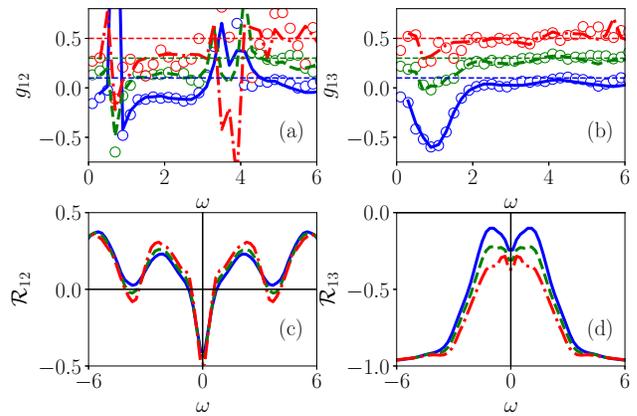}
	\caption{FDT indicator functions $g_{12}(\omega)$ in (a) and
		$g_{13}(\omega)$ in (b) for $L=24$ and
	$\beta_\textrm{T}$ = 0.1~(solid), 0.3~(dashed), 0.5~(dashed dotted).
Also shown are the finite size correction form with symbols. The overlap
functions $\mathcal{R}_{12}(\omega)$ and $\mathcal{R}_{13}(\omega)$ 
are shown in (c) and (d) with the same parameter values.}
	\label{fig:4}
\end{figure}

We also perform the analysis for two-operators cases with
$\op{A} = \op{O}_1$ and $\op{B} = \op{O}_2$ or 
$\op{O}_3 \equiv \sum_l (\op\sigma_l^+ \op\sigma_{l+1}^- + 
\op\sigma_l^- \op\sigma_{l+1}^+)$~[kinetic energy].
We evaluate the FDT indicator functions
at the energy eigenstates with $\beta_\textrm{T}=0.1$, 0.3, and 0.5. 
The numerical results for the largest
system size $L=24$ are presented in Fig.~\ref{fig:4}(a) and (b). The indicator
functions exhibit intermittent fluctuations and seem to deviate from 
$\beta_\textrm{T}$ significantly. We will show that the apparent deviations
observed in Figs.~\ref{fig:2} and \ref{fig:3} are indeed the finite size 
effect.

Our theory predicts that the indicator function should suffer from a finite
size effect described by \eqref{beta_fsc}. In the energy eigenstate with
$\Delta^2E=0$, only the intrinsic term $\mathcal{Y}^{(1)}$ 
contributes to the finite size effect and the  
deviation from the FDT is given by~\cite{Note1}
\begin{equation}\label{d_beta_intrinsic}
	\delta\beta_{AB}(\omega) = \frac{\partial}{\partial E} \ln\left[
f_{A}(E,-\omega) f_B(E,\omega)\mathcal{R}_{AB}(E,\omega)\right] .
\end{equation}
When $\op{A}=\op{B}$, the overlap function $\mathcal{R}_{AA}(E,\omega)=1$ 
and the correction term becomes
\begin{equation}
	\delta\beta_{AA}(\omega) = \frac{\partial}{\partial E}\ln \left[
  f_A(\bar{E},-\omega)f_A(\bar{E},\omega)\right] .
  \label{delta_beta_AA}
\end{equation}
The function $f_A(E,\omega)$ determining the fluctuation amplitude 
of offdiagonal matrix elements in the ETH can be evaluated numerically. 
We explain our numerical method in Supplemental Material~\cite{Note1}. 
For $L=24$, we evaluate $\delta\beta$ in \eqref{delta_beta_AA}
numerically, and compare the indicator function $g$ and thus-obtained
$\beta+\delta\beta$ in Fig.~\ref{fig:2}. 
The two curves $g(\omega)$ and 
$\beta+\delta\beta$ are in good agreement. 

We also test the finite size effect for $\op{A}\neq\op{B}$.
The overlap functions are evaluated at the energy 
values corresponding to the inverse temperature $\beta=0.1$, 0.3, and
0.5~\cite{Note1}. They are plotted in Fig.~\ref{fig:4}(c) and (d).
Using the numerical data, we can evaluate the finite size correction term in
\eqref{d_beta_intrinsic}. Figure~\ref{fig:4} (a) and (b) show that the
indicator function and the finite size correction theory are in excellent
agreement. The overlap function $\mathcal{R}_{12}(\omega)$ 
has zeros, at which the correlation function $\bar{S}_{12}(\omega) \propto
\mathcal{R}_{12}(\omega)$ also vanishes. The intermittent fluctuations in 
$g_{12}(\omega)$ occur near the zeros.  
$\mathcal{R}_{13}(E,\omega)$ varies more rapidly with $\beta$ at small
values of $\omega$, explains the strong finite size
effect for $g_{13}(\omega)$.
We have also investigated the FDT, the finite size effect, and the overlap
functions for five different observables~\cite{Note1}. We add a remark that
the numerical results do not depend on a particular choice of
$\Delta\omega$~(see Fig. S4 in \cite{Note1}).

The function $f$ and $\mathcal{R}$ may scale with the system size 
$L$~\cite{LeBlond:2019bv, Mierzejewski:2020jm}. 
An overall scale factor, if any, cancels out 
in taking the logarithmic derivative in \eqref{d_beta_intrinsic}. 
Because the derivative is taken with
respective to the extensive quantity $E=O(L^d)$, the correction term 
scales as $\delta\beta_{AB}(\omega) = O(L^{-d})$.
The inset of Fig.~\ref{fig:3}(b) confirms that numerical data
are consistent with the finite size scaling 
$(\beta_{\mathrm{FDT}}-\beta_{\mathrm T})=O(L^{-1})$.
Therefore, we conclude that our numerical data confirms the FDT 
in the infinite size limit.

{\em Discussions and summary---}
The FDT plays crucial roles in various fields of condensed matter physics, since it 
can be used to extract information on the response to an
external perturbation from equilibrium fluctuations. Our result can
serve as a theoretical foundation of the FDT for pure quantum states beyond
the conventional setup with the Gibbs states.   This is particularly
relevant to ultracold atoms, for which the FDT has been experimentally
investigated~\cite{Gemelke:2009ja,Meineke:2012bv}.  For example, in
Ref.~\cite{Meineke:2012bv}, the magnetic susceptibility
is obtained from measurements of equilibrium fluctuations.  We expect that
the FDT is also experimentally useful for temperature measurements of
isolated systems, as is numerically confirmed in the present work. 

Verifying the FDT experimentally in isolated quantum systems is a challenge. 
The KMS condition, explained in this work, can be tested 
in an experiment with a frequency-resolved measurement of correlation 
functions. 
Suppose that the energy variance is negligible.
Combining \eqref{S_AB_result} and \eqref{d_beta_intrinsic} and eliminating
the microscopic overlap function, 
the finite size correction term $\delta\beta_{AB}$ 
can be rewritten as
\begin{equation}
	\delta\beta_{AB} = -\frac{\hbar\omega}{2}\frac{\partial
	\beta}{\partial E} + \frac{\partial}{\partial E} \ln \bar{S}_{AB}
	+ O(L^{-2d}) ,
\end{equation}
which involves the quantities experimentally accessible.
Our theory for the finite size correction will be useful in an
experimental study because experimental system sizes available are definitely
finite~\cite{Kaufman:2016jm,Gross:2017do}.

\begin{acknowledgments}
This work is supported by the National Research Foundation of Korea (NRF)
grant funded by the Korea government (MSIP) (Grants No. 2019R1A2C1009628~(JDN)
and No. R2017R1D1A09000527~(JY)).
TS is supported by JSPS KAKENHI Grant Numbers JP16H02211 and JP19H05796.  
TS is grateful to Takeshi Fukuhara and Shuta Nakajima for valuable discussions.
\end{acknowledgments}

\bibliography{paper}

\renewcommand{\theequation}{S\arabic{equation}}
\renewcommand{\thefigure}{S\arabic{figure}}
\setcounter{equation}{0}
\setcounter{figure}{0}
\widetext
\newpage
\begin{center}
{\bf\large Supplemental Materials}\\
\vspace{4mm}
\setcounter{page}{1}

Jae Dong Noh${}^{1}$, Takahiro Sagawa${}^{2}$, and Joonhyun Yeo${}^{3}$\\
\vspace{2mm}

${}^{1}${\it Department of Physics, University of Seoul, Seoul 02504,
Korea}\\

${}^{2}${\it Department of Applied Physics, The University of Tokyo,\\
7-3-1 Hongo, Byunkyo-ku, Tokyo 113-8656, Japan} \\

${}^{3}${\it Department of Physics, Konkuk University, Seoul 05029, Korea}
\end{center}

\section{Derivation of the KMS condition Eq.~(\ref{KMS_w}) from the ETH}

Applying the ETH to the operators $\op{A}$ and $\op{B}$, one can write 
\begin{equation}\label{app:Sab1}
\bar{S}_{AB}(\omega) = 2\pi \sum_{\alpha} p_\alpha 
    \sum_{\gamma\neq\alpha} 
    e^{-S(E_{\gamma\alpha})}
    f_A(E_{\alpha\gamma},\omega_{\alpha\gamma})
    f_B(E_{\gamma\alpha},\omega_{\gamma\alpha})
    R^A_{\alpha\gamma} R^B_{\gamma\alpha} 
    \delta(\omega-\omega_{\gamma\alpha}) .
\end{equation}
The energy eigenvalues are densely distributed for large system sizes. 
Thus, one can replaces $\sum_\gamma$ with $\int dE_\gamma D(E_\gamma)$ 
with the density of state function $D(E_\gamma) = e^{S(E_\gamma)}$. 
The individual matrix elements of $R^A$ and $R^B$ are random variables
with zero mean and unit variance~\cite{Srednicki:1996kn,DAlessio:2016gr}. 
On the other hand, $R^A_{\alpha\gamma}$ and $R^B_{\gamma\alpha}$ sharing 
the same eigenstates may be correlated with a nonvanishing value of
$R^A_{\alpha\gamma}R^B_{\gamma\alpha}$ on average. Along the line 
of the ETH, we make an ansatz that $R^A_{\alpha\gamma}R^B_{\gamma\alpha}$
in \eqref{app:Sab1} can be replaced by a smooth function 
$\mathcal{R}_{AB}(E_{\gamma\alpha},\omega_{\gamma\alpha})$,
which will be called the overlap function. It satisfies
\begin{equation}
    \mathcal{R}_{AB}(E,\omega) = \mathcal{R}_{AB}(E,-\omega)^* = 
    \mathcal{R}_{BA}(E,-\omega)
\end{equation}
for Hermitian operators $\op{A}$ and $\op{B}$. When $\op{A}=\op{B}$, 
$\mathcal{R}_{AA}(E_{\gamma\alpha},\omega_{\gamma\alpha}) = 
2 \delta_{\alpha\gamma} + (1-\delta_{\alpha\gamma})$.
Then, the correlation function can be written as
\begin{equation}\label{Sab_cont}
    \bar{S}_{AB}(\omega) = 2\pi \sum_\alpha p_\alpha 
    e^{-S(E_\alpha+\hbar\omega/2) + S(E_\alpha+\hbar\omega)}
	Q_{AB}(E_\alpha+\hbar\omega/2,w) 
\end{equation}
with the auxiliary function
\begin{equation}
    Q_{AB}(E,\omega) \equiv f_A(E,-\omega) f_B(E,\omega)
	\mathcal{R}_{AB}(E,\omega) = Q_{AB}(E,-\omega)^* = Q_{BA}(E,-\omega).
\end{equation}

We proceed with small $\delta E_\alpha = (E_\alpha - \bar{E}) = O(L^{d/2})$
expansion with the mean energy $\bar{E} =  {\rm Tr} \rho_i \op{H} = O(L^d)$.
The entropy term becomes
\begin{equation}
 \begin{split}
    S(E_\alpha+\hbar\omega) - S(E_\alpha+\hbar\omega/2) =& 
    S(\bar{E} + \delta E_\alpha + \hbar\omega)
    - S(\bar{E} + \delta E_\alpha +\hbar\omega/2) \\
    =&S(\bar{E} + \hbar\omega)
    - S(\bar{E} +\hbar\omega/2) +  (\delta E_\alpha)  \{ \partial_E S(\bar{E} + \hbar\omega) -\partial_E S(\bar{E} + \hbar\omega/2) \} \\
    &+\frac 1 2 (\delta E_\alpha)^2 \{ \partial^2_E S(\bar{E} + \hbar\omega) -\partial^2_E S(\bar{E} + \hbar\omega/2) \} +
     O\left( \frac{(\delta E_\alpha)^3}{\bar{E}^3} \right).
 \end{split}
\end{equation}
On the other hand, we have
\begin{equation}  \begin{split} \label{Q_expand} 
    \ln Q_{AB}(E_\alpha+\hbar\omega/2,\omega) = & \ln Q_{AB}(\bar{E}+\hbar\omega/2,\omega) 
     + (\delta E_\alpha) \partial_E \ln Q_{AB}(\bar{E}+\hbar\omega/2,\omega) \\
    &+\frac{1}{2} (\delta E_\alpha)^2 \partial_{E}^2 \ln Q_{AB}(\bar{E}+\hbar\omega/2,\omega)
    +O\left( \frac{(\delta E_\alpha)^3}{\bar{E}^3} \right) .
    \end{split} 
\end{equation}
We put the above two equations into Eq.~(\ref{Sab_cont}) and expand the exponential in powers of
$\delta E_\alpha$. We then reexponentiate it after averaging over the initial distribution given by $p_\alpha$
to obtain
\begin{equation} \label{S_AB_1}
\begin{split}
\bar{S}_{AB}(\omega)= 2\pi \exp &\Big[ S(\bar{E} + \hbar\omega)
    - S(\bar{E} +\hbar\omega/2)+\ln Q_{AB}(\bar{E}+\hbar\omega/2,\omega) \\
    &+ \frac 1 2 \Delta^2 E \Big\{ \partial_E S(\bar{E} + \hbar\omega) -\partial_E S(\bar{E} + \hbar\omega/2)
    + \partial_{E}\ln Q_{AB}(\bar{E}+\hbar\omega/2,\omega) \Big\}^2 \\
    & +\frac 1 2 \Delta^2 E \Big\{ \partial^2_E S(\bar{E} + \hbar\omega) -\partial^2_E S(\bar{E} + \hbar\omega/2)
    + \partial_{E}^2 \ln Q_{AB}(\bar{E}+\hbar\omega/2,\omega) \Big\}+O\left( \frac{\overline{(\delta E_\alpha)^3}}{\bar{E}^3}\right)\Big]   ,
    \end{split}
\end{equation}
where $\Delta^2 E \equiv \sum_\alpha p_\alpha (\delta E_\alpha)^2$ is the
energy variance of the initial state.

Now we note that $\beta = (\partial_{\bar{E}} S)$ is the inverse temperature at
the energy $\bar{E}$ in 
the microcanonical ensemble, which is an intensive quantity. If we expand the quantities in Eq.~(\ref{S_AB_1}) around $\bar{E}$,
each derivative with respect to $\bar{E}$ contributes a factor of $O(L^{-d})$. We can therefore write
\begin{equation} \label{S_AB_final}
\begin{split}
\bar{S}_{AB}(\omega)=& 2\pi \exp \left[ \frac{\hbar\omega}{2}\beta+\ln Q_{AB}(\bar{E},\omega)
 +\left\{ \frac{3(\hbar\omega)^2}{8} (\partial_{\bar{E}}\beta)  +
 \frac{\hbar\omega}{2}\partial_E \ln Q_{AB}(\bar{E},\omega)\right\} +
 O\left(\frac{1}{\bar{E}^2}\right) \right. \\
 & \left. +\frac 1 2 \Delta^2 E \Big\{  
    \Big(\frac{\hbar\omega}{2}\partial_{\bar{E}} \beta 
    + \partial_{E} \ln Q_{AB}(\bar{E},\omega) \Big)^2 
    +\frac{\hbar\omega}{2}\partial^2_{\bar{E}} \beta 
    + \partial_{E}^2 \ln Q_{AB}(\bar{E},\omega)+O\left(
    \frac{1}{\bar{E}^3}\right) \Big\}+O\left( \frac{\overline{(\delta
    E_\alpha)^3}}{\bar{E}^3}\right) \right]   \\
    =&  2\pi e^{\frac{1}{2}\beta\hbar\omega} 
  Q_{AB}(\bar{E},\omega)  \exp\left[ \mathcal{Y}_{AB}(\bar{E},\omega) 
  + O\left(\max\left\{\frac{1}{\bar{E}^{2}},  \frac{\Delta^2
  E}{\bar{E}^3},\frac{\overline{(\delta E_\alpha)^3}}{\bar{E}^3}\right\}\right)  \right] ,
    \end{split}
\end{equation}
where $\mathcal{Y}_{AB}(E,\omega) = \mathcal{Y}^{(1)}_{AB}(E,\omega) +
\mathcal{Y}^{(2)}_{AB}(E,\omega)$ with  
\begin{align}
	\mathcal{Y}^{(1)}_{AB}(E,\omega) =&  \frac{3(\hbar\omega)^2}{8} (\partial_E\beta) +
	\frac{\hbar\omega}{2} \frac{\partial}{\partial E} \ln Q_{AB}(E,\omega), 
 \label{C1_res} \\
 \mathcal{Y}^{(2)}_{AB}(E,\omega) =& \frac 1 2 \Delta^2 E \left[ 
 \frac{(\hbar\omega)^2}{4}(\partial_E\beta)^2+
 \frac{\hbar\omega}{2}\partial^2_{E} \beta  +\hbar\omega (\partial_E\beta)(\partial_E\ln Q_{AB}(E,\omega))
    + \frac{ \partial_{E}^2 Q_{AB}(E,\omega)}{Q_{AB}(E,\omega)}
 \right]   . \label{C2_res}
\end{align}

Notice that $\mathcal{Y}^{(1)}_{AB}$ involves the partial derivative of the
scale-independent quantities with respect to the extensive quantity $E=O(L^d)$. 
Thus, it scales as $\mathcal{Y}^{(1)} = O(L^{-d})$.
The second term $\mathcal{Y}^{(2)}_{AB}$ is nonzero only when the initial state has an energy uncertainty 
with nonzero $\Delta^2 E$. In addition, it involves the partial derivative
with respect to the energy twice. Thus, it scales as $\mathcal{Y}^{(2)}_{AB}
= O(\Delta^2 E/ E^2)$.
When $\Delta^2 E = O(L^d)$ as in ordinary noncritical thermal systems, 
$\mathcal{Y}^{(2)}_{AB}(E,\omega) = O(L^{-d})$.
Therefore, in the infinite system size limit, the correction term
$\mathcal{Y}$ vanishes and the correlation function becomes 
$\bar{S}_{AB}(\omega) = 2\pi e^{\frac{1}{2}\beta\hbar\omega}
Q_{AB}(\bar{E},\omega)$, which obeys the KMS condition.

\section{Numerical method for $\bar{S}_{AB}$ and $\mathcal{R}_{AB}$}
In this section, we explain the numerical method to evaluate the correlation
function $\bar{S}_{AB}$ and the overlap functions $\mathcal{R}_{AB}$.
As a prerequisite, we assume that the complete set of energy eigenstates $\{ \ket{\alpha}\}$ and 
the matrix elements for $\op{A}$ and $\op{B}$ in the energy eigenstate basis
are ready. 

We will evaluate the correlation function defined in \eqref{S_AB_formal} at discrete values of 
$\omega_n = (n+1/2) \Delta\omega$ with $n=0,\pm 1,\pm 2,\cdots$. 
It is given by $\bar{S}_{AB}(\omega_n) =
\frac{1}{\Delta\omega}\int_{\omega_n-\Delta\omega/2}^{\omega_n+\Delta\omega/2}
d\omega \bar{S}_{AB}(\omega)$, and can be evaluated as
\begin{equation}
  \bar{S}_{AB}(\omega_n) = \frac{1}{\Delta\omega} \sum_\alpha p_\alpha
  \left( \sum_{E_\alpha-\Delta\omega/2 < E_\gamma < E_\alpha+\Delta\omega/2}
  A_{\alpha\gamma} B_{\gamma\alpha} \right) .
\end{equation}
If the system is in an energy eigenstate represented by $\op{\rho}_i = \ket{\alpha_0}
\bra{\alpha_0}$, then $p_\alpha = \delta_{\alpha \alpha_0}$. In the main text,
we investigate the correlation function for the energy eigenstate. In order to
reduce fluctuations, we choose $p_\alpha = \mbox{constant}$ for
$|E_\alpha-E_{\alpha_0}| < \Delta\omega/2$ and $p_\alpha = 0$ otherwise.

\begin{figure}[t]
  \includegraphics[width=0.5\columnwidth]{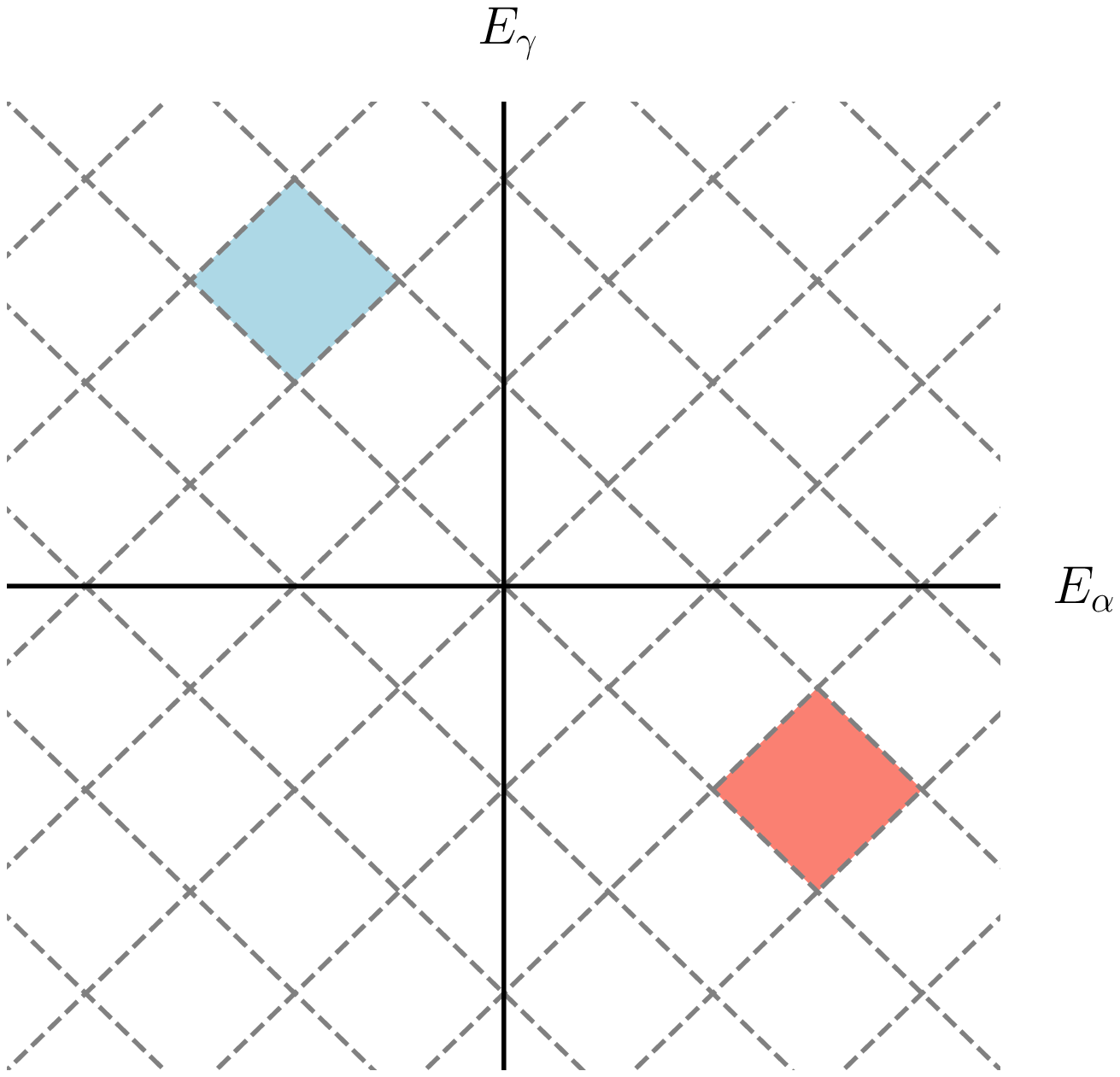}
  \caption{$(E_\alpha,E_\gamma)$ plane with meshes for discrete values of $E =
  (E_\gamma+E_\alpha)/2$ and $\omega=(E_\gamma-E_\alpha)/2$ in unites of $\Delta\omega$.
  Each cell is represented by the coordinate $(E,\omega)$ at the central point. 
  The two shaded cells share the same value of $E$ and have the opposite values of $\omega$.}
  \label{fig:S1}
\end{figure}

The functions $f_A(E,\omega)$ and $\mathcal{R}_{AB}(E,\omega)$ for $\omega\neq
0$ determine the statistical properties of offdiagonal matrix elements of
observables in the energy eigenstate basis in the context of the ETH. 
We explain our method to evaluate those functions at discrete values of $E$
and $\omega$ in units of $\Delta\omega$. 
We first construct the table $D(E_n)$ for the density of states
by counting the number of energy levels
$\ket{\alpha}$'s within the interval $E_n-\Delta\omega/2 \leq  E_\alpha < E_n + \Delta\omega/2$.
It is related to the microcanonical ensemble entropy through $D(E_n) =
e^{-S(E_n)}$. Then, we separate all pairs of energy eigenstates into discrete sets, each
of which is characterized by $(E_n,\omega_m)$ and consists of pairs of 
eigenstates satisfying
\begin{equation}\label{cell}
	E_n - \frac{\Delta\omega}{2} \leq E_{\gamma\alpha}  \leq E_n + \frac{\Delta\omega}{2} \mbox{ and }
	\omega_m - \frac{\Delta\omega}{2} \leq \omega_{\gamma\alpha}  < \omega_m +
\frac{\Delta\omega}{2} .
\end{equation}
It may be helpful to imagine a two-dimensional $(E_\alpha,E_\gamma)$ plane as shown in Fig.~\ref{fig:S1}.
All pairs of eigenstates characterized by \eqref{cell} lie within a cell,
which will be denoted as $c_{n,m}$. We remind the readers that
$E_{\gamma\alpha} \equiv (E_\gamma+E_\alpha)/2$ and $\omega_{\gamma\alpha}
\equiv E_\gamma-E_\alpha$.

According to the ETH, an offdiagonal elements of an observable $\op{A}$ is
given by $A_{\gamma\alpha} = e^{-S(E_{\gamma\alpha})/2}
f_A(E_{\gamma\alpha},\omega_{\gamma\alpha}) R^A_{\gamma\alpha}$ where 
$R^A_{\gamma\alpha}=(R^A_{\alpha\gamma})^*$ has the same statistical property as the Gaussian
random variable with zero mean and unit variance~\cite{Srednicki:1996kn}.
Using the statistical property of $R^A$, one can isolate the amplitude of
$f_A$ by calculating
\begin{equation}
	|f_A(E_n,\omega_m)|^2 = \frac{1}{|c_{n,m}|}
	\sum_{(E_\alpha,E_\gamma)\in c_{n,m}} D(E_n) A_{\gamma\alpha}
	A_{\alpha\gamma} ,
	\label{fA_numerics}
\end{equation}
where $|c_{n,m}|$ is the number of pairs within cell. The factor $D(E_n)$
cancels the entropy factor.
The overlap function
$\mathcal{R}_{AB}(E_{\gamma\alpha},\omega_{\gamma\alpha})$ for
$R^A_{\alpha\gamma} R^B_{\gamma\alpha}$ can be also constructed by
calculating
\begin{equation}
	\mathcal{R}_{AB}(E_n,\omega_m) = \frac{1}{|c_{n,m}|}
	\sum_{(E_\alpha,E_\gamma)\in c_{n,m}} \frac{D(E_n)}{\sqrt{
		|f_A(E_n,-\omega_m)|^2 |f_B(E_n,\omega_m)|^2}}
	A_{\alpha\gamma}B_{\gamma\alpha}
	\label{R_AB_numerics}
\end{equation}
Note that $f_A(E,-\omega) = f_A(E,\omega)^*$ and
$\mathcal{R}_{AB}(E,-\omega) = \mathcal{R}_{AB}(E,\omega)^*$ for Hermitian
operators.

\begin{figure}[t]
	\includegraphics[width=\columnwidth]{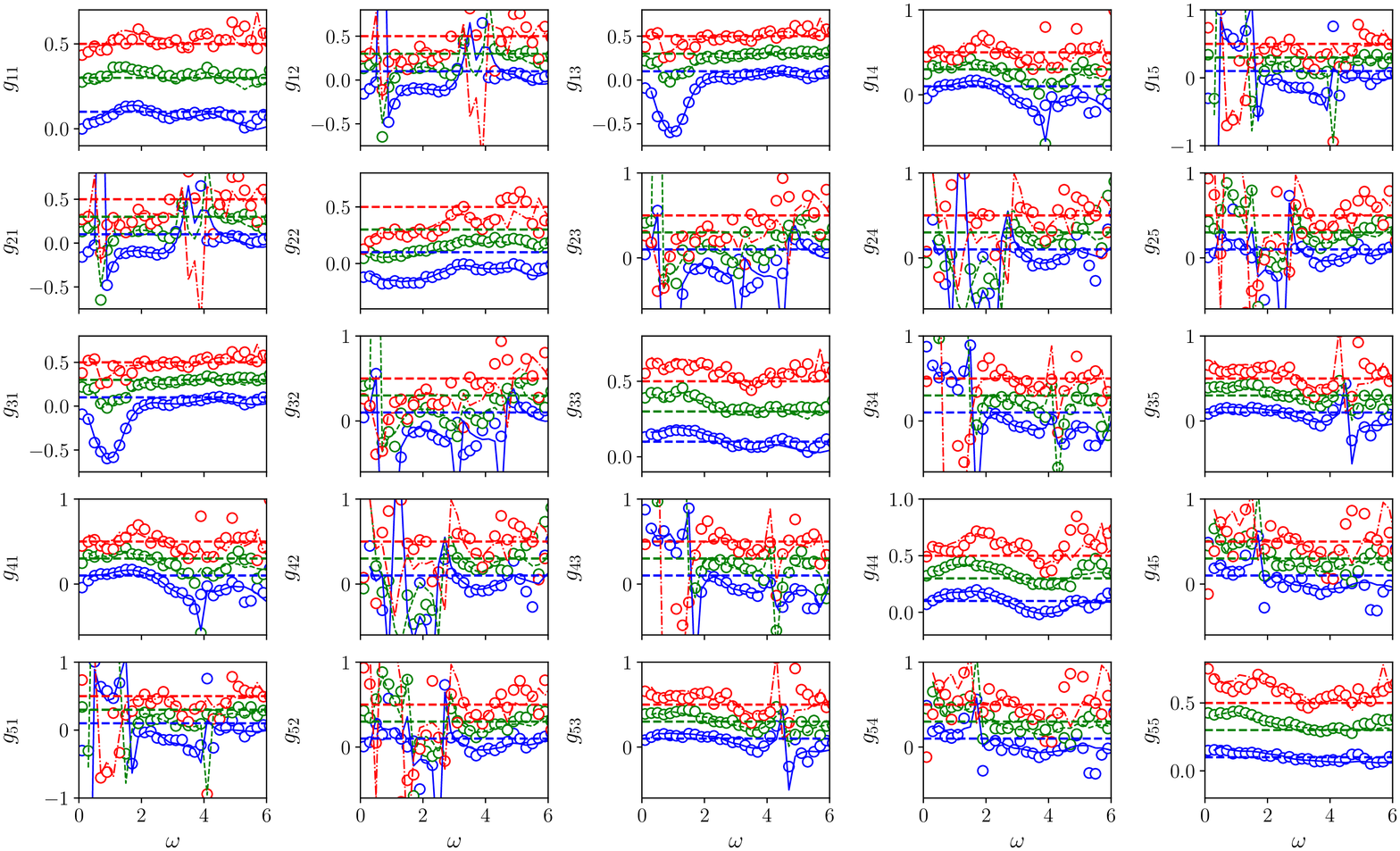}
	\caption{FDT indicator functions $g_{ij}(\omega)$~(lines) and their finite size
	correction form~(symbols) for the target inverse temperature $\beta=$
  0.1~(solid), 0.3~(dashed), and 0.5~(dashed dotted).}
	\label{fig:S2}
\end{figure}

\begin{figure}[t]
	\includegraphics[width=\columnwidth]{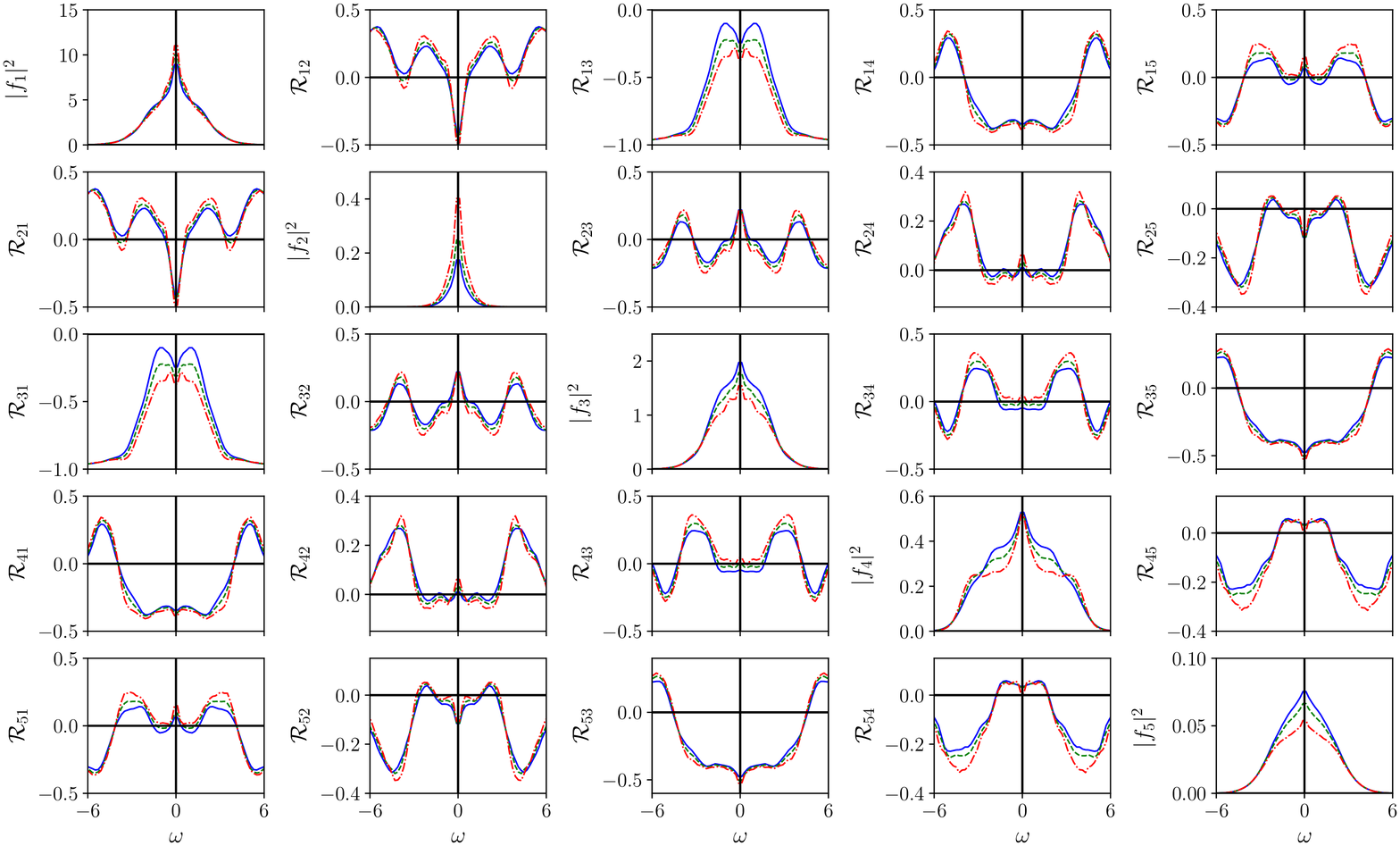}
	\caption{$|f_i(E,\omega)|^2$ and $\mathcal{R}_{ij}(E,\omega)$ as a
	  function of $\omega$ for the operators $\op{O}_i$ in \eqref{O_all}. 
	  The curves are evaluated at the energy value $E$ corresponding to the
	  inverse temperature $\beta=0.1$~(solid), 0.3~(dashed), and 0.5~(dashed
	  dotted). The diagonal and offdiagonal panels show $|f_i|^2$ and
  $\mathcal{R}_{ij}$, respectively.}
	\label{fig:S3}
\end{figure}

In the numerical study for the XXZ spin chain, we have considered the five
different operators:
\begin{equation}\label{O_all}
\begin{aligned}
\op{O}_1 &= \sum_l \op{\sigma}_l^z\op{\sigma}_{l+1}^z \\
\op{O}_2 &= \frac{1}{L} \sum_{l,m} \op{\sigma}_l^+\op{\sigma}_{m}^- \\
\op{O}_3 &= \sum_l \left( \op{\sigma}_l^+ \op{\sigma}_{l+1}^- +
	\op{\sigma}_l^-\op{\sigma}_{l+1}^+ \right) \\
	\op{O}_4 &= \frac{1}{L} \sum_{l,m} (-1)^{(l-m)}
	\op{\sigma}_l^z \op{\sigma}_m^z \\
	\op{O}_5 &= \frac{1}{L} \sum_{l,m}
	(-1)^{(l-m)}\op{\sigma}_l^+\op{\sigma}_{m}^- . \\
\end{aligned}
\end{equation}
There operators are Hermitian and even under the time reversal. Thus,
$f_A$ and $\mathcal{R}_{AB}$ are real valued functions.
The three operators $\op{O}_1$, $\op{O}_2$, and $\op{O}_3$ are considered  
in the main text. 
We evaluate the correlation functions $\bar{S}_{AB}$ for the operators
$\op{A}, \op{B} = \op{O}_{i}$ using
the method explained above, and then calculated the indicator functions
$g_{ij}(\omega)$. We take the energy eigenstate $\ket{\alpha_\textrm{T}}$ as
the initial state whose inverse temperature is closest to the target values
$\beta_\textrm{T}= 0.1$, $0.3$, and $0.5$. In order to reduce a statistical
fluctuation, we also perform the calculations for the initial states $\ket{\alpha}$ 
within the energy interval $E_\textrm{T} -\Delta\omega/2 < E_\alpha < E_\textrm{T} +
\Delta\omega/2$, and take the average over them.  
All the numerical data at system size $L=24$ are plotted in Fig.~\ref{fig:S2}
along with the finite size correction form $\beta = \beta+\delta\beta_{ij}$.
In Fig.~\ref{fig:S3}, we also present the plot of $|f_i(\omega)|^2$ and
$\mathcal{R}_{ij}(E,\omega)$ as a function of $\omega$ at the energy values
corresponding to $\beta = 0.1$, 0.3, and 0.5.

\begin{figure}[t]
	\includegraphics[width=\columnwidth]{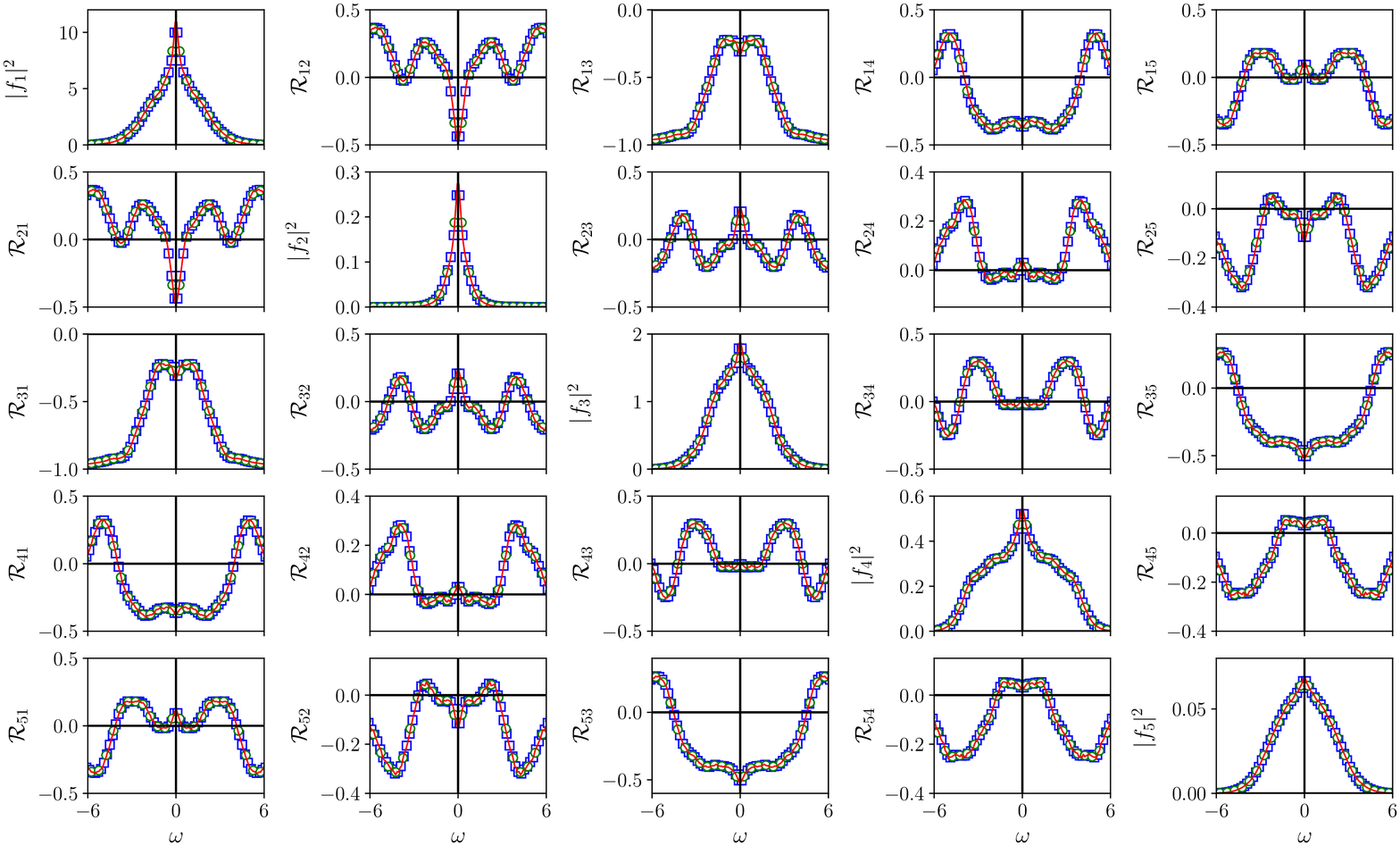}
    \caption{The same plots as in Fig.~\ref{fig:S3} with $\beta=0.3$. The
    data are obtained with the energy
    discretization $\Delta \omega = 0.1$~(line), $0.2$~(square symbol), and
  $0.5$~(circular symbol).}
	\label{fig:S4}
\end{figure}

The results do not depend on the coarse-graining scheme with $\Delta \omega < 
1.0$. In Fig.~\ref{fig:S4}, we compare the numerical data obtained with
$\Delta\omega=0.1$, $0.2$, and $0.5$ for the system of size $L=24$ and of
inverse temperature $\beta=0.3$. All the data sets are hardly
distinguishable. The comparison demonstrates that the existence of the
overlap function is not an artifact of the coarse-graining with a finite
value of $\Delta \omega$.

\end{document}